\documentclass[amssymb,twocolumn,aps]{revtex4}
\usepackage{graphics,color,graphicx,amsmath}
\usepackage{subcaption}
\usepackage{natbib}
\usepackage{verbatim}
\usepackage{graphicx}
\usepackage[above,below]{placeins}
\usepackage{float}
\usepackage{tabularx}
\usepackage{times}
\usepackage{blindtext}
\usepackage{url}
\usepackage{rotating}

\renewenvironment{quote}
  {\list{}{\rightmargin=.1cm \leftmargin=.3cm}%
   \item\relax}
  {\endlist}

\begin{document}

\title{Introductory physics students' recognition of strong peers: Gender and racial/ethnic bias differ by course level and context}

\author{Meagan Sundstrom,$^1$ Ashley B. Heim,$^2$ Barum Park,$^3$ and N. G. Holmes$^1$}
\affiliation{$^1$Laboratory of Atomic and Solid State Physics, Cornell University, Ithaca, New York 14853, USA\\
$^2$Department of Ecology and Evolutionary Biology, Cornell University, Ithaca, New York 14853, USA\\
$^3$Department of Sociology, Cornell University, Ithaca, New York 14853, USA}

\date{\today}

\begin{abstract}
    Researchers have pinpointed recognition from others as one of the most important dimensions of students' science and engineering identity. Studies, however, have found gender biases in students' recognition of their peers, with inconsistent  patterns across introductory science and engineering courses. Toward finding the source of this variation, we examine whether a gender bias exists in students' nominations of strong peers across three different remote, introductory physics courses with varying student populations (varying demographics, majors, and course levels). We also uniquely evaluate possible racial/ethnic biases and probe the relationship between instructional context (whether lecture or laboratory) and recognition. Some of our results replicate previous findings (such as the the association of course grade and small class section enrollment with nominations), while others offer contradictions. Comparing across our three courses and the prior work, results suggest that course level (whether first-year students or beyond-first-year students) might be more associated with a gender bias in peer recognition than other variables. Surprisingly, we also find instances of racial/ethnic biases in favor of students from backgrounds historically underrepresented in science. Finally, we find that the nomination patterns differ when students nominate individuals strong in the lecture material versus laboratory material. This work serves as an important step in determining which courses and contexts exhibit biases in peer recognition, as well as how students' perceptions of one another form in remote teaching environments.
\end{abstract}

\maketitle

\section{Introduction}

Introductory science, technology, engineering, and math (STEM) courses are particularly critical transition periods for students. These courses impact students' retention and persistence in their subsequent courses, their majors, and their future careers~\cite{suresh2006relationship,watkins2013retaining}. Retention rates, however, are lowest for students from marginalized groups, including women and underrepresented racial minorities (URM; taken as any racial/ethnic group other than White and Asian/Asian American from here on)~\cite{beasley2012they,riegle2019does,clark2005,sax2016,cheryan2017,merner2015african,white2018race,johnson2020women,porter2019women}. Research suggests that these retention issues are in part due to societal stereotypes that position women and URM students as less suitable to scientific fields than men and non-URM individuals~\cite{danielsson2012exploring,gonsalves2016masculinities,kessels2006goes,schmader2004,makarova2015trapped,makarova2019,carlone2007understanding,tate2005does,ceglie2011underrepresentation,grossman2014perceived,blaine2013understanding}. This positioning leads underrepresented students to experience low sense of belonging and low self-efficacy in their science and engineering courses~\cite{hall1982, gunter2005,seymour1997,rosenthal2011,cwik2021damage,eddy2014,moss2012,tate2005does}. These stereotypes manifest in many ways, including students' perceptions of their peers' abilities related to the course material~\cite{grunspan2016, salehi2019, bloodhart2020}. 

Prior work has found conflicting results with regard to whether there is a gender bias in introductory STEM students' perceptions of their peers~\cite{grunspan2016,salehi2019,bloodhart2020}. These studies observed that students disproportionately nominate men over women as strong in their biology and physics courses~\cite{grunspan2016,bloodhart2020}, but that men and women receive comparable numbers of nominations in mechanical engineering courses~\cite{salehi2019}. Whether these discrepant results are due to varying student populations (e.g., students' majors, academic years, and demographics) in the observed courses, the scientific discipline of the course, or some other factor is still unresolved. We advance this body of work by collecting and analyzing students' nominations of strong peers in three different remote physics courses serving various student populations. 
We aim to determine how the student population of and context within a course (lecture or laboratory) are related to students' perceptions of their peers. We also expand previous work by considering possible racial/ethnic biases, in addition to gender biases, in students' nominations. 

\vspace{-0.2cm}

\subsection{Recognition in STEM}

We situate our study in the theoretical framework of identity. An individual's identity refers to their being a ``certain kind of person" in a given context~\cite{gee2000chapter}. \textit{Science identity}, therefore, is the degree to which an individual believes they are a ``science person." Researchers have conceptualized a model of science and engineering identity containing four dimensions: performance, competence, interest, and recognition~\cite{carlone2007understanding,hazari2010connecting}. 
Studies show that \textit{recognition} is one of the most important of these dimensions in predicting students' participation, persistence, and career intentions in science and engineering~\cite{hyater2018critical,lock2013physics,carlone2007understanding,hazari2010connecting,hazari2018towards,hazari2017importance,kalender2019gendered,godwin2016identity,godwin2017pushing}. Recognition is the degree to which meaningful others (e.g., peers, teachers, and family) perceive and acknowledge an individual as a science person. When a student receives ample recognition from others, they are likely to see themselves as a science person and develop a strong science identity~\cite{hazari2018towards, kalender2019female}.


Given the importance of recognition, it stands a reasonable desire for all students to feel recognized by their peers in a science classroom. Recognition, however, is ``culturally produced": it is influenced by sociohistorical norms and stereotypes~\cite{carlone2007understanding,kalender2019gendered,avraamidou2022identities}. For physics in particular, stereotypes often position men and non-URM students as more suitable to the field than women and URM students~\cite{danielsson2012exploring,gonsalves2016masculinities,kessels2006goes,schmader2004,makarova2015trapped,makarova2019,carlone2007understanding,tate2005does,ceglie2011underrepresentation,grossman2014perceived,blaine2013understanding,moss2012,eddy2014}. In turn, empirical work shows that men and non-URM students report higher senses of recognition in their physics classes than women and URM students, respectively~\cite{lock2013physics,kalender2019gendered,hazari2013science}. Close examinations of women and women of color in physics~\cite{archer2017exceptional,carlone2007understanding,avraamidou2022identities} also affirm that high-achieving or ``exceptional" women in the field hinge on recognition from others to succeed. Thus, students' gender \textit{and} race/ethnicity are closely related to their recognition and identity in physics. Previous work, however, has not analyzed racial/ethnic bias in students' nominations of strong peers (they analyze gender bias only)~\cite{grunspan2016,salehi2019,bloodhart2020}. To contribute to this gap, we measure both gender and racial/ethnic bias in the current study.

Prior research also suggests that recognition may vary within different disciplines, instructional formats, and student populations. Grunspan and colleagues~\cite{grunspan2016} examined three iterations of a large, introductory biology course (the second in the course sequence) with gender-balanced enrollment. They found in all semesters that men received significantly more nominations as strong in the course material than women. The extent to which this bias occurred, however, varied between instruction types: they observed that women nominated other women more frequently when the instructor employed `random call' during class. Salehi and colleagues~\cite{salehi2019} later performed similar work for two offerings (one with traditional instruction and one with active learning instruction) of a medium-sized, gender-balanced mechanical engineering course taken by second and third-year students planning to major in engineering. They expected that the nature of the engineering discipline would lead to more gender bias in peer recognition than that found in biology. Instead, they found no gender bias in their observed nominations in either course offering. Lastly, Bloodhart and colleagues~\cite{bloodhart2020} analyzed peer perceptions across many introductory life science and physics courses. They do not report the student populations of each course, but, in aggregate, students in the life science courses were mostly non-URM women in their first year of study and the physical science courses were mostly non-URM men across all four academic years. The researchers found in the two disciplines that both men and women under-nominated women as knowledgeable in the course material in comparison to women's actual final grades in the class. To examine whether the different results across these three studies are attributable to varying disciplines and student populations, we analyze three physics courses serving different populations of students.

The studies above also suggest that instruction type, such as a traditionally taught lecture versus an active learning course with frequent group work, may influence recognition. In physics, furthermore, recognition likely varies between the instructional contexts of lecture and laboratory (lab) work. Not only do these two contexts involve very different pedagogies (lectures contain many students who focus on the instructor and labs contain a small number of students who collaborate on tasks), but they also cover distinct content and aim to develop different sets of skills~\cite{Phys21,kozminski2014aapt,holmes2018introductory,Smith2021,smith2020direct,holmes2015teaching}. Correspondingly, research has shown that students relate knowledge of mathematics or theoretical physics with lectures, while they view ``doing lab" as handling machinery and using technical skills~\cite{gonsalves2016masculinities,danielsson2012exploring}. Such differences in relevant skill sets suggest that different students may be recognized as strong in lecture and lab contexts. Thus, we probe and analyze students' recognition of peers in the two contexts separately in the current study.

\subsection{Current study}

We collected students' nominations of peers they believed were strong in lecture and lab contexts in three different remote, introductory physics courses at Cornell University. These data allowed us to compare peer recognition across instructional contexts (within our study) as well as across disciplines (comparing our study of physics courses to prior research on courses in other disciplines). Each of the three physics courses also contained varying student populations in terms of students' major, academic year, gender, and race/ethnicity. This variation allowed us to examine whether gender and racial/ethnic biases in peer recognition depend on features of student population as suggested by prior studies~\cite{grunspan2016,salehi2019,bloodhart2020}. The following two research questions guided our study:

\begin{enumerate}
\itemsep0em 
    \item To what extent do gender and racial/ethnic biases exist in students' recognition of strong peers across three different introductory physics courses serving distinct student populations?
    \item How do introductory physics students’ recognition of strong peers differ in lecture and lab contexts?
\end{enumerate}

Comparing the three courses in our study to previous studies~\cite{grunspan2016,salehi2019,bloodhart2020}, we find that whether students' perceptions of strong peers in lecture exhibit a gender bias might depend on course level over other variables (e.g., student populations or scientific discipline of the course). Courses serving first-year students exhibit a gender bias in lecture perceptions, while those serving the beyond-first-year level do not. Surprisingly, we also find in some cases that URM students tend to receive more nominations than their non-URM peers. With regard to instructional context, we observe that both the general patterns of nominations and whether gender or racial/ethnic bias exists in nominations differ between the lecture and lab contexts.

\section{Methods}

In this section, we first characterize the courses and students we analyzed. Then, we describe our data collection and analysis methods.

\begin{table*}[t]
\caption{\label{tab:demographics}%
Summary of instruction modality, student demographics, and student final course grades for the three courses we analyzed. All online meetings were synchronous. Percentages are relative to the number of students included in the analysis. 
We denote students' gender or race/ethnicity as `unknown' if they preferred not to disclose this information on the survey or if they did not complete the survey (but did consent to other parts of the research). Grades are provided on a 4.0 scale.
}
\begin{ruledtabular}
\setlength{\extrarowheight}{1pt}
\begin{tabular}{lccc}
\textrm{}&
\textrm{Course A}&
\textrm{Course B}&
\textrm{Course C}\\
\colrule
Modality  \\
\hspace{5mm}Lecture & Online & Online & Online \\
\hspace{5mm}Lab sections & Online & Online & Online \\
\hspace{5mm}Discussion sections & 12 Online, 2 In-person & 3 Online, 2 In-person & Online \\
Total enrollment & 208 & 89 & 433 \\
Students in analysis  & 198  & 84  & 404  \\
Gender \\
\hspace{5mm}Men & 84 (42\%) & 59 (70\%) & 182 (45\%) \\
\hspace{5mm}Women & 92 (47\%) & 23 (28\%) & 206 (51\%) \\
\hspace{5mm}Non-binary & 0 & 1 (1\%)  & 3 (1\%) \\
\hspace{5mm}Unknown & 22 (11\%) & 1 (1\%)  & 13 (3\%) \\

 Race/ethnicity\\
\hspace{5mm}Non-URM & 140 (71\%) & 68 (81\%)  & 268 (66\%) \\
\hspace{5mm}URM & 32 (16\%) & 12 (14\%)  & 107 (27\%) \\
\hspace{5mm}Unknown race/ethnicity & 26 (13\%) & 4 (5\%) & 29 (7\%) \\
 
Major\\
\hspace{5mm}Physics/Engineering Physics & 11 (5\%)  & 58 (69\%) & 14 (3\%) \\
\hspace{5mm}Engineering  & 128 (65\%)  & 14 (17\%)  & 331 (83\%)  \\
\hspace{5mm}Other STEM  & 19 (10\%) & 7 (8\%) & 24 (6\%)\\
\hspace{5mm}Unknown  & 40 (20\%)  & 5 (6\%)  & 35 (8\%) \\

Year \\
\hspace{5mm}First-year & 166 (83\%)& 78 (93\%)  & 10 (2\%)\\
\hspace{5mm}Second-year  & 24 (12\%) & 3 (4\%)  & 374 (93\%)  \\
\hspace{5mm}Third-year  & 6 (3\%)  & 1 (1\%)  & 12 (3\%) \\
\hspace{5mm}Other/Unknown & 2 (2\%) & 2 (2\%) & 8 (2\%)  \\

Grades \\
\hspace{5mm}Mean (SD) final course grade (men) & 3.6 (0.6) & 3.5 (0.6) & 3.4 (0.5)\\
\hspace{5mm}Mean (SD) final course grade (women) & 3.6 (0.4) & 3.2 (0.6) & 3.4 (0.4) \\
\hspace{5mm}Mean (SD) final course grade (non-URM) & 3.5 (0.6) & 3.5 (0.6) & 3.4 (0.4)\\
\hspace{5mm}Mean (SD) final course grade (URM) & 3.6 (0.6) & 3.2 (0.5) & 3.2 (0.5) \\
\end{tabular}
\end{ruledtabular}
\end{table*}

\subsection{Courses and participants}

Our data come from three introductory physics courses at Cornell University, which we call Courses A, B, and C. These courses were held during the Fall 2020 semester, when about three-quarters of students at the institution resided on campus but most courses were held online due to the COVID-19 pandemic. Course A was an introductory, calculus-based mechanics course aimed at first-year students intending to major in engineering or other STEM disciplines. Course B was also an introductory, calculus-based mechanics course, but primarily served first-year students intending to major in physics. Course C was an introductory, calculus-based electromagnetism course intended for first- and second-year students majoring in engineering or other STEM disciplines. Course C is typically taken after Course A in a course sequence, so it is reasonable to assume that students were familiar with one another to some extent before entering the course. Students in our data set for Courses A and C are different, however, because we examine the two courses during one semester. Only data for Course C provide a snapshot of this familiarity as many of those students took mechanics during the previous semester, but that offering of Course A is not analyzed here.

As summarized in Table \ref{tab:demographics}, each course had lecture, lab, and discussion sections. Most course components were held online (synchronously), with a few held in person. Lectures for all three courses were instructed by a male faculty member in the physics department. Courses A and C were ``flipped," such that students read relevant sections of the textbook and took a reading quiz before coming to class. Lectures for Course A used conceptual iClicker questions and instructor demonstrations, while lecture time in Course C was spent on problem-solving questions through Learning Catalytics~\cite{newland2021review}. In both courses, students answered questions both individually and in groups. Course B was more traditional in that, during lectures, the instructor presented new content and asked iClicker questions that students answered individually. 
Courses B and C (but not Course A) used an online discussion forum where both students and teaching staff could post questions and answers related to course content at any time. 

\begin{table*}[t] 
\centering
\caption{\label{tab:meannoms}
Survey response rates and mean number of nominations made per student. The survey response rate is the percent of enrolled students who completed the survey. The mean number of nominations is the average number of peers' names that a student listed.}
\begin{ruledtabular}
\setlength{\extrarowheight}{1pt}
\begin{tabular}{lccc}
& Course A & Course B & Course C\\
\hline
Survey response rate & 79\% & 92\% & 83\%\\
Mean nominations (lecture) & 1.2 & 1.9 & 1.0 \\ 
Mean nominations (lab) & 1.2 & 1.3 & 0.9 \\ 
\end{tabular}
\end{ruledtabular}
\end{table*}

In all three courses, lab and discussion sections were led by graduate teaching assistants. Lab sections met once per week for two hours and discussion sections met twice per week for 50 minutes. Each section contained approximately 20 students who worked together in small groups of two to four. The labs and most of the discussion sections took place online through Zoom and students worked in groups in virtual breakout rooms. In the few discussion sections held in person, students worked together at round tables. Labs were inquiry-based~\cite{Phys21,kozminski2014aapt,holmes2018introductory,Smith2021,smith2020direct,holmes2015teaching}, where students performed open-ended investigations using objects at home or in their dorm rooms. They submitted their lab work as a group, rather than individually. During discussion sections, students solved problems related to the lecture content. Despite working as a group, students submitted their discussion work individually in Course C. Students were not required to submit discussion work in Courses A and B.

We collected students' self-reported gender, race/ethnicity, intended major, and academic year via a survey at the beginning of the semester. We grouped race/ethnicity by URM status, where non-URM students are those solely identifying as White and/or Asian/Asian American and URM students are those identifying as at least one of any other race/ethnicity (including Black or African American, Hispanic/Latinx, and Native Hawaiian or other Pacific Islander). These student populations and average final course grades are summarized in Table \ref{tab:demographics}. Similar proportions of men (42\%) and women (47\%) were enrolled in Course A and roughly three-quarters of the students in this course identified as non-URM  (71\%). Men and women received comparable final course grades on average in this course, as did URM and non-URM students. Most students in Course B were men (70\%) and more than three-quarters of the students identified as non-URM (81\%). Men and non-URM students on average received higher final course grades in this course than women and URM students, respectively. Course C contained similar proportions of men (45\%) and women (51\%) and two-thirds of the students identified as non-URM (66\%). Men and women on average received comparable final course grades in this course, while non-URM students on average received higher final course grades than URM students.

\subsection{Data collection}

In all three courses, we administered an online survey during the eighth week of the 15-week semester as part of a lab assignment about students' group work experiences. On the survey, we asked students to nominate peers who they believed were knowledgeable in the course with the following two prompts adapted from prior work~\cite{grunspan2016,salehi2019,bloodhart2020}:
\begin{quote}
    \textit{Please list any students in this physics class that you think are particularly strong in the lecture/discussion section material.}\\
    \\
    \textit{Please list any students in this physics class that you think are particularly strong in the lab material.}
\end{quote}
We refer to the first prompt as ``lecture perceptions" and the second prompt as ``lab perceptions."

The survey was in an open response format (one text box) and students could respond with an unlimited number of names. This format avoids students feeling obligated to fill a quota and writing down extra names of peers they may not actually perceive as strong~\cite{grunspan2014}. Students were also not given a class roster from which to choose or look up names. This resulted in some listings being hard to match to the class roster during analysis, as there were instances of students misspelling peers' names and reporting just a first or a last name. Thus, text processing of the responses was necessary. We compared each name reported on the survey to the class roster and matched up names for which the number of corrections needed to match the full name on the roster was fewer than 0.3 times the length of someone's full name. We chose the constant 0.3 via trial and error, finding that this worked best for capturing as many close matches as possible without producing false negatives. If a name (either first or last) appeared multiple times in the data set, then we did not match on listings of just that name itself and only matched listings of the other half of the name or the full name.



As summarized in Table \ref{tab:meannoms}, survey response rates were high (all above 75\%) and students in each course on average listed one or two peers for each prompt. Our analysis included all students who responded to the survey and/or were listed by at least one peer. We also only included the nominations made by students who consented to participate in research (more than 95\% of survey responders). If a consenting student nominated a non-consenting student, we included the nomination, but removed all information (demographics, etc.) about the non-consenting student. In all courses, at least 93\% of the enrolled students are included in our analysis (see Table \ref{tab:demographics}). 

We note that prior studies~\cite{grunspan2016,salehi2019} used surveys late in the semester, which formed highly centralized networks (many nominations were concentrated to a few students) for course-level perceptions. In our study, two of the three courses exhibited highly centralized lecture perception networks at the mid-semester mark, so there is no reason to believe our results are impacted by the timing of survey administration.


At the end of the semester, we collected discussion and lab section enrollment for all courses. For courses B and C, we also collected the number of student contributions to the course's online discussion forum (sum of their posted questions and posted answers to others' questions). Course A did not use a discussion forum. We used these discussion forum contributions as a measure of students' outspokenness because it quantifies students' communicative engagement during an online course. This is a similar, but distinct, measure to that of Grunspan and colleagues~\cite{grunspan2016}, who determined outspokenness by asking the instructor to name actively participating students after each class meeting. Some students had no discussion forum data, indicating that they likely did not ever register for or use the forum. For these students, we imputed their contributions to the discussion forum as zero, which was also the mode of each course's distribution of contributions. We imputed one student's and 55 students' discussion forum data in Courses B and C, respectively.

\subsection{Analysis of nominations}

We converted the survey responses into directed networks for each course (A, B, and C) and each context (lecture and lab). \textit{Nodes} represented students and \textit{edges} (or \textit{ties}) represented all nominations made between students. To first gain a sense of each network's overall structure, we calculated two network-level statistics: \textit{density} and \textit{indegree centralization}. Density measures the proportion of all possible edges in the network that we observed. Indegree centralization measures the extent to which the nominations are concentrated around a single student or a small subset of students (i.e., whether there are emergent \textit{celebrities} who receive most of the nominations). Higher indegree centralization indicates higher concentration around one or a few students. We determined the standard errors of each of these statistics via bootstrapping: resampling the observed network many times, calculating the statistic of each sampled network, and then determining the standard deviation of the statistic among all of the sampled networks~\cite{traxler2020network,snijders1999non}. The bootstrapping was performed with 10,000 bootstrap trials for each network using the \textit{snowboot} package in R~\cite{chen2019snowboot}. 


We then used exponential random graph models (ERGMs) to statistically determine the salient structural characteristics of our networks. ERGMs assume that the observed network is a realization from a random graph that comes from a distribution belonging to the exponential family~\cite{anderson1999,robins2007}. They allow us to perform many statistical tests at once, determining whether the frequency of certain patterns or configurations in our observed network is significantly different than if the ties were formed randomly. To formulate such a model, we first choose a principled set of predictor variables (i.e., configurations) that might explain the formation of the observed network. These variables may be structural (e.g., measuring the tendency for mutual nominations) or nodal (e.g., measuring the extent to which students of a certain gender are more likely to receive a nomination). 
The goal is to use these $k$ network statistics $g_k(y)$ and their corresponding coefficients $\theta_k$ to predict the formation of the random network $Y$. The model takes the form
\begin{equation*}
    P_\theta[Y = y] = \frac{1}{\psi}\exp\left(\sum_{k} \theta_k g_k(y)\right)
\end{equation*}
where $y$ is a realization of the random network $Y$ and $\psi = \sum_y \exp\left(\sum_{k}\theta_k g_k(y)\right)$ is a normalization constant that ensures that the probability sums to one. Given an observed network $y$, the coefficients of the model are estimated using Maximum Likelihood Estimation (MLE). Due to the dependence between the network ties, the MLE is commonly approximated with Markov Chain Monte Carlo (MCMC) techniques~\cite{hunter2008}, which we used to fit all models in our analysis.

There are two different ways to interpret the coefficients of ERGMs. In general, the coefficients weight the importance of each modeled configuration for the formation of the realized network, where positive (negative) coefficients show that the configuration is observed more (less) frequently than by chance after accounting for all other configurations that are modeled. The second way to interpret the coefficients is to focus on specific ties of the network. In this ``change statistics" interpretation, the coefficient $\theta_k$ of the $k$th configuration shows how much the log-odds of a tie being present changes if the formation of the tie increases the $k$th configuration by one unit, holding the rest of the network constant. For instance, if the predictor variable measures the number of mutual ties in the network, its coefficient represents how much the log-odds of a tie being present increases when the addition of this tie would reciprocate an existing tie.



We initially fit ERGMs with the same set of predictor variables used by Grunspan and colleagues~\cite{grunspan2016} for each of our observed networks. Our model contained one additional variable for discussion section homophily (the tendency for students to nominate peers enrolled in their same discussion section), because discussion was an extra structural component in our courses. We also added three variables to measure effects of race/ethnicity, which exactly mirrored the structure of the gender variables in the original model of Grunspan and colleagues~\cite{grunspan2016}. Inspection of the goodness-of-fit diagnostics, however, revealed a significant inadequacy in this model: we were not appropriately capturing the presence of \textit{triadic closure}. Triadic closure is the tendency for three nodes to be connected, given pairwise connections. That is, if ties exist between nodes A and B and between nodes B and C, then a tie between nodes A and C forms triadic closure. In some cases, we were also not adequately capturing the network's \textit{outdegree} distribution (the proportion of nodes \textit{making} a certain number of nominations). Our model did sufficiently account for each network's \textit{indegree} distribution (the proportion of nodes \textit{receiving} a certain number of nominations). 

In response, we altered the original model from Grunspan and colleagues~\cite{grunspan2016}. We added a geometrically-weighted edgewise shared partner (GWESP) variable, which is typically used to account for triadic closure. The more ties two nodes have in common (i.e., the more shared partners they have), however, the higher the probability of an edge forming between them. Thus, a decay parameter for the GWESP variable determines the extent to which the probability of tie formation decreases for each additional partner already shared between two nodes~\cite{hunter2007curved}. This parameter can take on a value between 0 and 1, with lower values creating larger decreases in tie probability per subsequent shared partner. We used a fixed decay parameter of 0.25 as is commonly used in ERGM literature~\cite{hummel2012improving,yin2021highly,butts2014introduction}. Because incorporating both a GWESP term and an isolates term (for students receiving zero nominations; used in the original model) produced degeneracy in the model, we removed the isolates term. We also changed the structure of the gender and race/ethnicity variables to allow for easier and more meaningful interpretations. Specifically, we added a variable to the model for each possible directed tie for the gender and race/ethnicity attributes (e.g., man nominating a man, man nominating a woman, etc.) as in Ref.~\cite{wells2019}. These network statistics allowed for a more direct comparison of ties by using common base terms for gender and race/ethnicity variables and thus an easier interpretation of gender and racial/ethnic biases. We note that in creating these variables, we fit models with each possible base term for gender and race/ethnicity to the observed networks. We ultimately chose to use nominations between majority demographic groups as the base terms, however the results are consistent with those of all the possible models.

These modifications resulted in an improved model fit for every observed network. For all six observed networks, the goodness-of-fit diagnostics showed that we were capturing the distributions of indegree, outdegree, and triadic closure well with the revised model. Coefficient estimates using the original model of Ref.~\cite{grunspan2016} and an example of goodness-of-fit metrics for both models are provided in the Appendix. We report in the main text our results using the revised model, which contained the following predictor variables:

\begin{enumerate}
    \itemsep0em
    \item \textit{Edges}: intercept term equal to the number of edges in the network
    \item \textit{Mutuality}: number of reciprocated or mutual nominations
    \item \textit{Geometrically weighted edgewise shared partners (GWESP)}: triadic closure
    \item \textit{Woman $\rightarrow$ woman}: number of edges for which a woman nominates another woman (base term is \textit{man $\rightarrow$ man})
    \item \textit{Woman $\rightarrow$ man}: number of edges for which a woman nominates a man (base term is \textit{man $\rightarrow$ man})
    \item \textit{Man $\rightarrow$ woman}: number of edges for which a man nominates a woman (base term is \textit{man $\rightarrow$ man})
    \item \textit{URM $\rightarrow$ URM}: number of edges for which a URM student nominates a URM student (base term is \textit{non-URM $\rightarrow$ non-URM})
    \item \textit{URM $\rightarrow$ non-URM}: number of edges for which a URM student nominates a non-URM student (base term is \textit{non-URM $\rightarrow$ non-URM})
    \item \textit{Non-URM $\rightarrow$ URM}: number of edges for which a non-URM student nominates a URM student (base term is \textit{non-URM $\rightarrow$ non-URM})
    \item \textit{Final course grade of nominee}: correlation between final course grade on indegree
    \item \textit{Discussion forum contributions of nominee} (only for Courses B and C): correlation between discussion forum contributions on indegree
    \item \textit{Homophily on lab section}: number of edges connecting students enrolled in the same lab section
    \item \textit{Homophily on discussion section}: number of edges connecting students enrolled in the same discussion section
\end{enumerate}
We determined the coefficient estimates of these variables for each of our six observed networks using MCMC MLE and then compared the results across courses and contexts.

\begin{figure*}[t]
\includegraphics[width=6in]{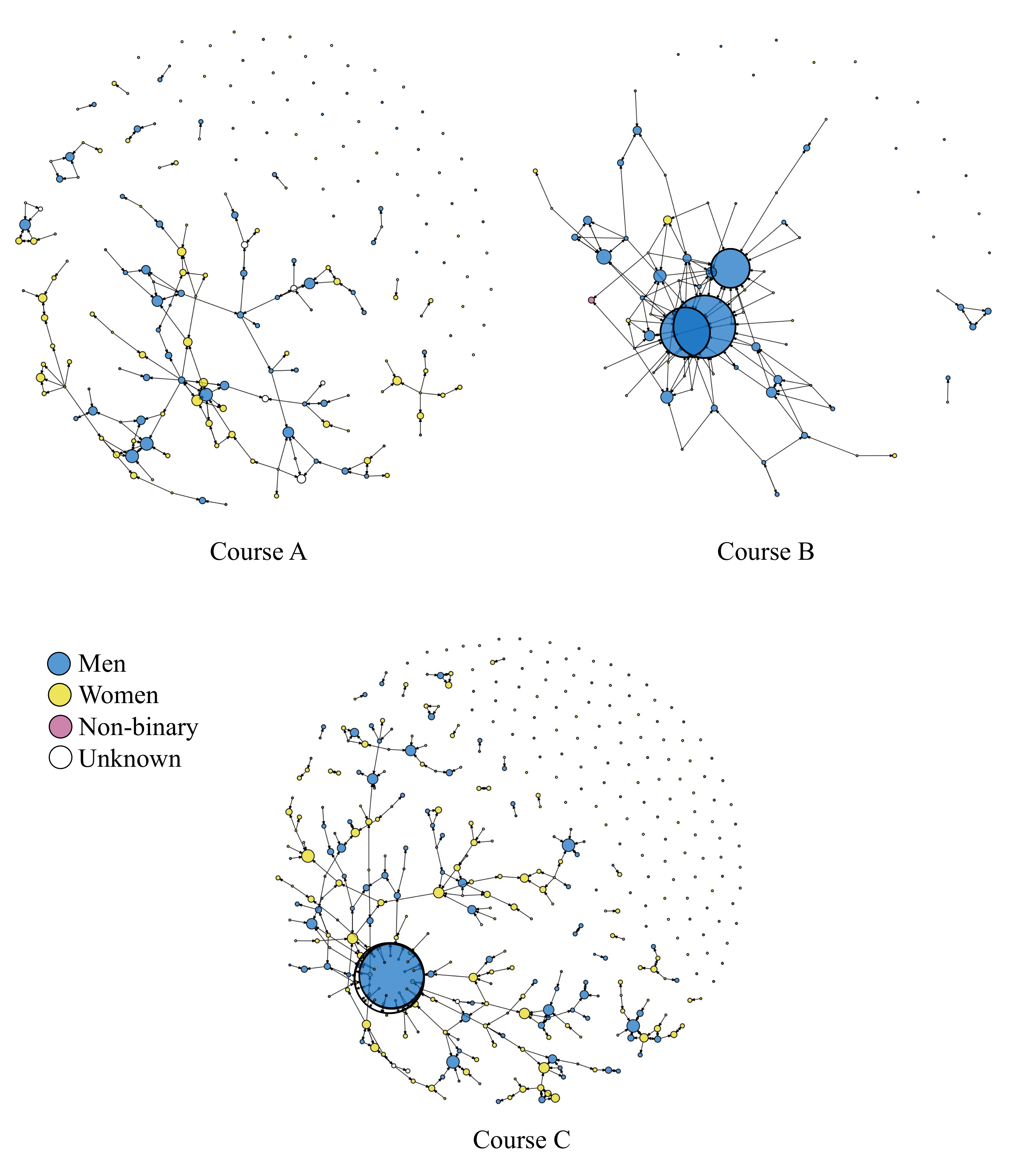}
\caption{Lecture perception networks. Nodes are colored by self-reported gender and sized proportional to indegree (number of received nominations). Nodes with bold outlines indicate celebrities (three in Course B and two in Course C). Edges point from the nominator to the nominee.}
\label{nonlabgendernetworks}
\end{figure*}

\begin{figure*}[t]
\includegraphics[width=6in]{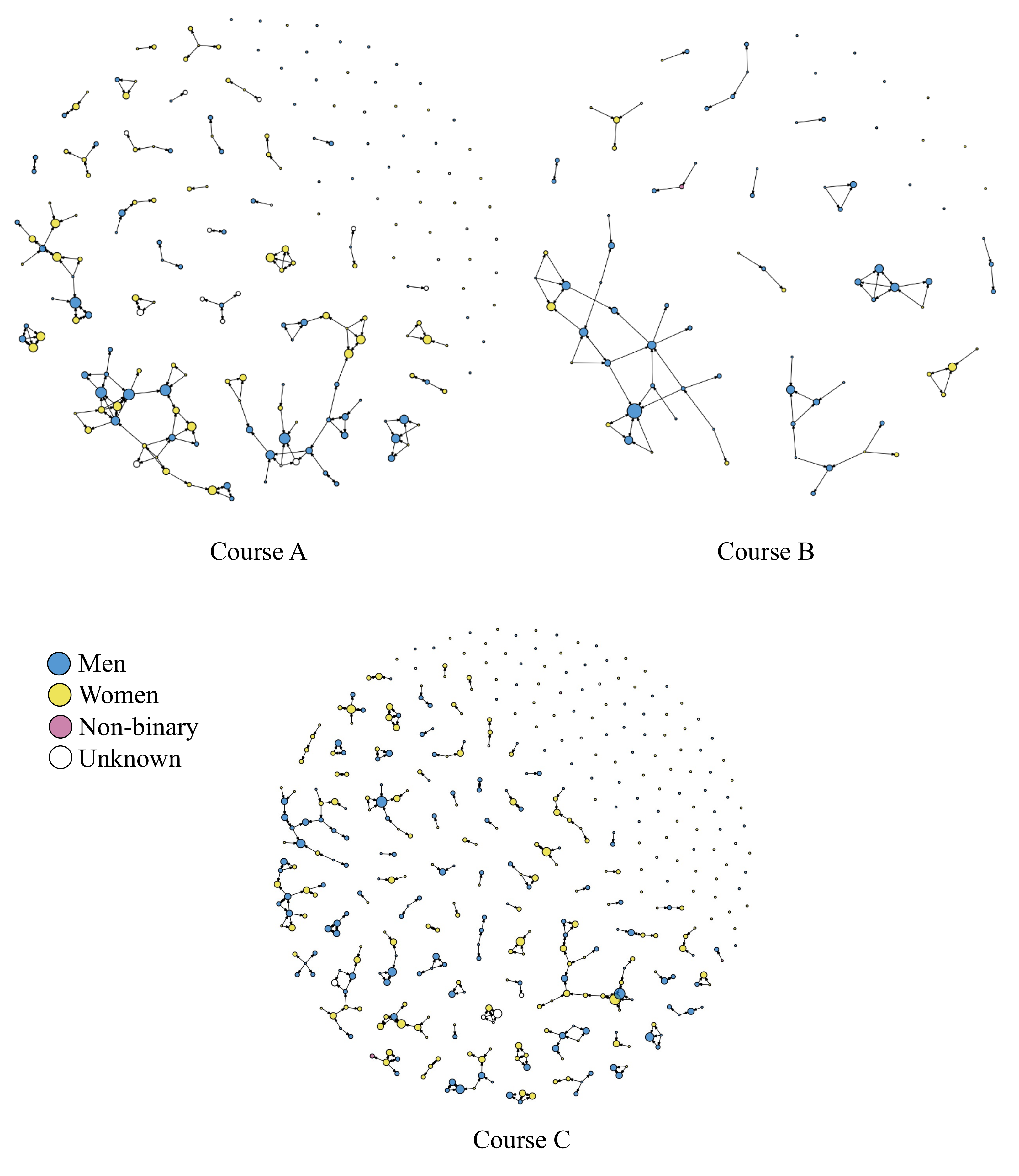}
\caption{Lab perception networks.
Nodes are colored by self-reported gender and sized proportional to indegree (number of received nominations). Edges point from the nominator to the nominee.}
\label{labgendernetworks}
\end{figure*}

\section{Results}

We first compare the structures and network-level statistics of each observed network. Then, we present the results of the exponential random graph models.

\subsection{Structure of peer perception networks}

Figures \ref{nonlabgendernetworks} and \ref{labgendernetworks} show the network diagrams of lecture and lab perception networks, respectively, for each course. In each diagram, edges point from the nominator to the nominee and larger nodes represent students who received more nominations (i.e., higher indegree). Nodes are colored by gender and nodes with bold outlines indicate celebrities. These same network diagrams with nodes colored by race/ethnicity can be found in the Supplementary Material.

Table \ref{tab:networkstats} summarizes the density and indegree centralization measures for each observed network. Within each course, the network densities of the lecture and lab perception networks are similar to one another. We see this as similar levels of connectedness (proportion of possible edges present) in Figs. \ref{nonlabgendernetworks} and \ref{labgendernetworks} for lecture and lab perception networks within each course. This similarity in network density, however, is despite the very different structures of these connections across contexts. That is, students nominate similar numbers of peers in each network, but the distribution of who receives the nominations is different between lecture and lab contexts.

\begin{table*}[t] 
\centering
\caption{\label{tab:networkstats}
Network-level statistics for all observed networks. Density is the proportion of observed to possible edges. Indegree centralization is the extent to which the nominations are concentrated on one or a few students. Standard errors of the last digit are shown in parentheses.}
\begin{ruledtabular}
\setlength{\extrarowheight}{1pt}
\begin{tabular}{lcccccc}
 & \multicolumn{2}{c}{Course A} &  \multicolumn{2}{c}{Course B} & \multicolumn{2}{c}{Course C} \\ 
 \cline{2-3}
 \cline{4-5}
 \cline{6-7}
 & Lecture & Lab & Lecture & Lab & Lecture & Lab \\ 
 \hline
 Density & 0.005(7) & 0.005(6) & 0.02(6) & 0.01(2) & 0.002(3) & 0.002(2) \\ 
 Indegree Centralization & 0.02(9) & 0.02(8) & 0.3(8) & 0.06(2) & 0.08(2) & 0.008(3) \\ 
\end{tabular} 
\end{ruledtabular}
\end{table*}

In Course A, the indegree centralization (the extent to which the network is concentrated around just a few students) is similarly low for both lecture and lab perceptions. Correspondingly, we see in Figs. \ref{nonlabgendernetworks} and \ref{labgendernetworks} that there are no emergent celebrities in either network for this course (no nodes are drastically larger than the others). In Courses B and C, however, the indegree centralization value for the lecture perception network is larger than that for the lab perception network by an order of magnitude. This suggests that the lecture perception networks of these two courses are much more concentrated around a few prominent students (celebrities) than the lab perception networks. We observe in Fig. \ref{nonlabgendernetworks} that the lecture perception networks of Courses B and C contain three and two celebrities (nodes with bold outlines that are much larger than the rest, having received many more nominations), respectively. On the other hand, we see in Fig. \ref{labgendernetworks} that there are no central nodes receiving many nominations in either of the two lab perception networks (all nodes are similar in size). Thus, for Courses B and C, despite the similar density measures, the lecture perception networks are much more concentrated around a few celebrities, with no outstanding celebrities in the lab perception networks.

\begin{table*}[!ht] 
\centering
  \caption{\label{tab:newmodel} Exponential random graph model results. Coefficient estimates of predictor variables for each observed network. We fit models with all possible permutations of the gender and race/ethnicity variables serving as the base terms, but here we show results using nominations between majority demographic groups as the base terms (\textit{man} $\rightarrow$ \textit{man} for gender and \textit{non-URM} $\rightarrow$ \textit{non-URM} for race/ethnicity). Standard errors of the coefficient estimates are in parentheses. Asterisks indicate statistical significance ($^{*}$p$<$0.05; $^{**}$p$<$0.01).} 
  \begin{ruledtabular}
\begin{tabular}{lcccccc} 
\\[-1.8ex] & \multicolumn{2}{c}{Course A} &  \multicolumn{2}{c}{Course B} & \multicolumn{2}{c}{Course C} \\ 
\cline{2-3}
\cline{4-5}
\cline{6-7}
\\[-1.8ex] & \multicolumn{1}{c}{Lecture} & \multicolumn{1}{c}{Lab} & \multicolumn{1}{c}{Lecture} & \multicolumn{1}{c}{Lab} & \multicolumn{1}{c}{Lecture} & \multicolumn{1}{c}{Lab} \\ 
\hline
 \textit{Edges} & -9.09$^{**}$ & -7.85$^{**}$ & -12.04$^{**}$ & -7.88$^{**}$ & -8.85$^{**}$ & -8.67$^{**}$ \\ 
  & (0.72) & (0.54) & (1.11) & (0.86) & (0.31) & (0.53) \\ 
  & & & & & & \\ 
 \textit{Mutuality} & 2.92$^{**}$ & 1.82$^{**}$ & 0.96$^{*}$ & 1.21$^{*}$ & 3.30$^{**}$ & 1.90$^{**}$ \\ 
  & (0.40) & (0.35) & (0.45) & (0.51) & (0.19) & (0.31) \\ 
  & & & & & & \\ 
  \textit{GWESP (triadic closure;} & 1.21$^{**}$ & 1.06$^{**}$ & 1.05$^{**}$ & 0.96$^{**}$ & 1.91$^{**}$ & 0.98$^{**}$ \\ 
 \textit{decay parameter = 0.25)} & (0.17) & (0.14) & (0.15) & (0.19) & (0.07) & (0.11) \\ 
  & & & & & & \\ 
 
   \textit{Woman $\rightarrow$ woman} & 0.06 & 0.08 & -0.70 & 0.47 & 0.46$^{***}$ & 0.01 \\ 
  & (0.17) & (0.15) & (0.73) & (0.28) & (0.09) & (0.18) \\ 
  & & & & & & \\ 
\textit{Woman $\rightarrow$ man} & -0.03 & -0.04 & -0.13 & -0.16 & 0.22 & -0.16 \\ 
  & (0.21) & (0.21) & (0.19) & (0.28) & (0.14) & (0.18) \\ 
  & & & & & & \\ 
 \textit{Man $\rightarrow$ woman} & -0.52$^{*}$ & -0.23 & -1.45$^{**}$ & -0.77$^{**}$ & 0.07 & -0.31 \\ 
  & (0.24) & (0.24) & (0.55) & (0.37) & (0.14) & (0.19) \\ 
  & & & & & & \\ 
 \textit{URM $\rightarrow$ URM} & -0.17 & 0.11 & 1.73$^{**}$ & 1.08$^{*}$ & 0.23 & -0.19 \\ 
  & (0.55) & (0.43) & (0.43) & (0.54) & (0.13) & (0.29) \\ 
  & & & & & & \\ 
\textit{URM $\rightarrow$ non-URM} & 0.34 & 0.24 & 0.31 & -0.04 & -0.40$^{**}$ & -0.24 \\ 
  & (0.24) & (0.28) & (0.22) & (0.33) & (0.14) & (0.19) \\ 
  & & & & & & \\ 
 \textit{Non-URM $\rightarrow$ URM} & -0.02 & -0.52 & -0.07 & -0.22 & 0.53$^{**}$ & -0.16 \\ 
  & (0.27) & (0.34) & (0.40) & (0.38) & (0.11) & (0.17) \\ 
  & & & & & & \\ 
  \textit{Final course grade of nominee} & 0.86$^{**}$ & 0.29$^{*}$ & 1.91$^{**}$ & 0.46$^{*}$ & 0.54$^{**}$ & 0.29$^{*}$ \\ 
  & (0.19) & (0.14) & (0.28) & (0.22) & (0.08) & (0.15) \\ 
  & & & & & & \\ 
 \textit{Discussion forum contributions} & N/A & N/A & 0.02$^{**}$ & 0.008 & 0.02$^{**}$ & -0.003 \\ 
 \textit{of nominee} &  &  & (0.003) & (0.006) & (0.005) & (0.01) \\ 
  & & & & & & \\ 
 \textit{Homophily on lab section} & 1.01$^{**}$ & 3.37$^{**}$  & 0.59$^{**}$ & 2.97$^{**}$ & 1.01$^{**}$ & 4.19$^{**}$ \\ 
  &  (0.17) & (0.21)  & (0.17) & (0.32) & (0.12) & (0.16) \\ 
  & & & & & & \\ 
 \textit{Homophily on discussion section} &  1.59$^{**}$ & 0.20 & 0.66$^{**}$ & 0.09 & 1.76$^{**}$ & 0.27 \\ 
  &  (0.17) & (0.25)  & (0.18) & (0.24) & (0.09) & (0.20) \\ 
  & & & & & & \\ 
  \hline
\end{tabular} 
\end{ruledtabular}
\end{table*} 


\begin{figure*}[t]
\includegraphics[width=6.5in,trim={0 2cm 0 1cm}]{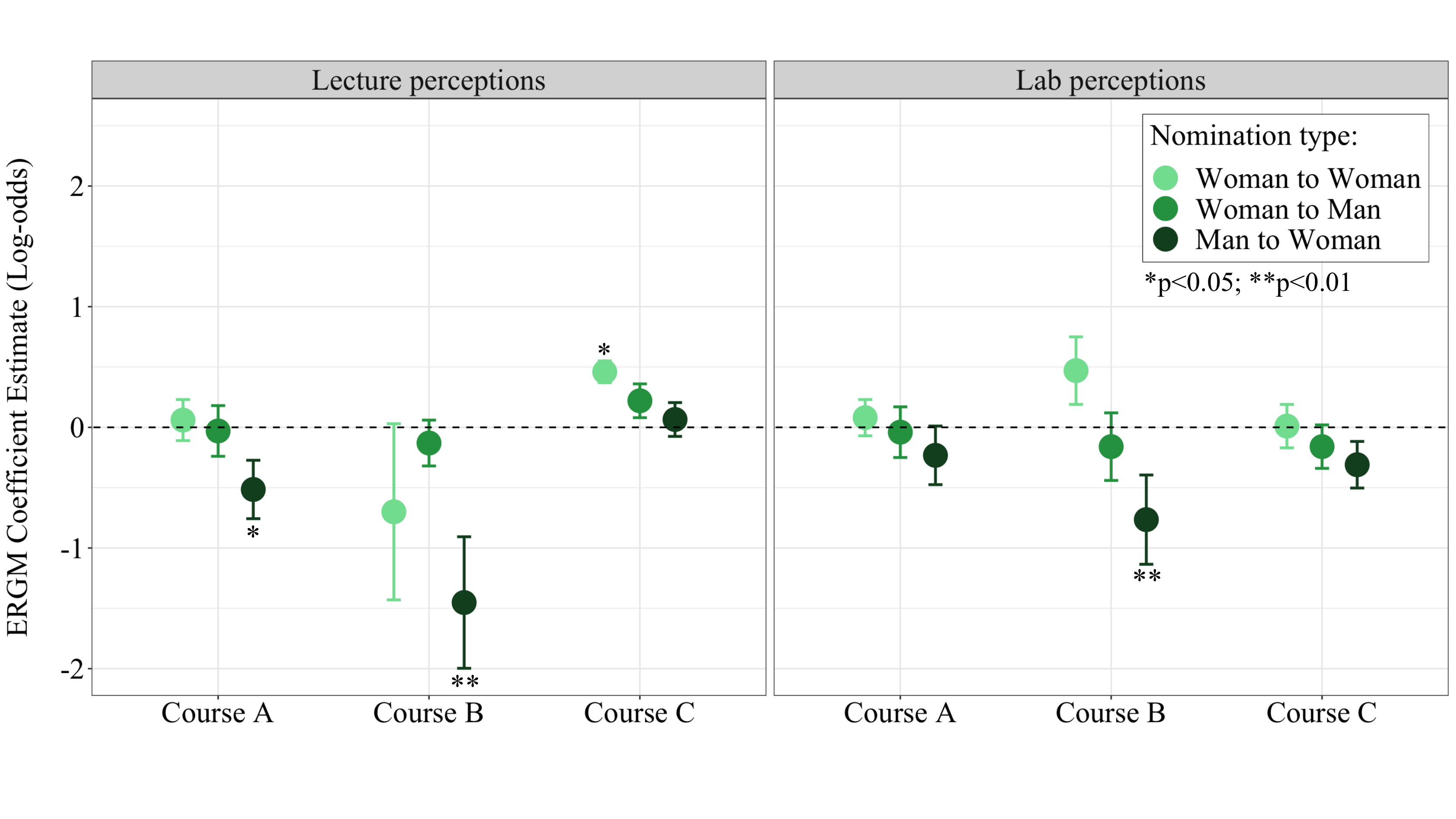}
\caption{Plot of ERGM coefficient estimates for the gender nomination variables. The base term (i.e., coefficient estimate of zero) is nominations from man to man. Error bars are the standard errors of the coefficient estimates (values shown in Table \ref{tab:newmodel}) and asterisks indicate statistical significance. 
}
\label{maletofemale}
\end{figure*}


\subsection{Evaluating gender and racial/ethnic bias in peer recognition}

Table \ref{tab:newmodel} shows the coefficient estimates for our revised exponential random graph model fit to all observed networks. We interpret the coefficient estimates as the log-odds of tie formation. For example, the coefficient estimate for the \textit{homophily on discussion section} variable for Course A's lecture perception network is 1.59. This means that the log-odds of a tie forming in the network increases by 1.59 for each additional tie connecting students in the same discussion section, holding the rest of the network the same. In other words, ties connecting students in the same discussion section are more probable than ties connecting students in different discussion sections, even after accounting for the other configurations included in the model.

For five out of six analyzed networks, the coefficient estimates for the \textit{woman $\rightarrow$ woman} and \textit{woman $\rightarrow$ man} variables, shown in light and medium green dots in Fig. \ref{maletofemale}, are not statistically significant. This means the frequency with which women nominate either a woman or man is not significantly different than the frequency with which men nominate other men (the base term) after adjusting for the other variables in the model. In other words, women proportionately nominate their female and male peers in these five networks. In the lecture perception network of Course C, however, women nominate other women significantly more than men nominate other men. 


The coefficient estimates for the \textit{man $\rightarrow$ woman} variable, shown in dark green dots in Fig. \ref{maletofemale}, indicate that men significantly under-nominate women in the lecture perception network of Course A and both networks of Course B. The lecture and lab perception networks of Course B, moreover, have the largest and second-largest coefficient magnitudes for this variable, respectively. This suggests that the strongest gender bias occurs in Course B's lecture perception network. Making direct comparisons of ERGM coefficients across different networks, however, has limitations~\cite{duxbury2021problem}, so we consider this claim preliminary.

We note that the coefficient estimate for the \textit{man $\rightarrow$ woman} variable is not statistically significant in the lecture perception network of course C, however this might be due to one of the two celebrities having unknown gender. If we impute this celebrity's gender as a man, the \textit{man $\rightarrow$ woman} variable becomes negative and statistically significant (implying a gender bias against women), though the other gender variables are not statistically significant. If we impute this celebrity's gender as a woman, the results related to gender are the same as when this celebrity's gender is unknown. The dependency of the statistical results on this one celebrity's gender offers an important caveat to our interpretations discussed in the next section.

\begin{figure*}[t]
\includegraphics[width=6.5in,trim={0 2cm 0 1cm}]{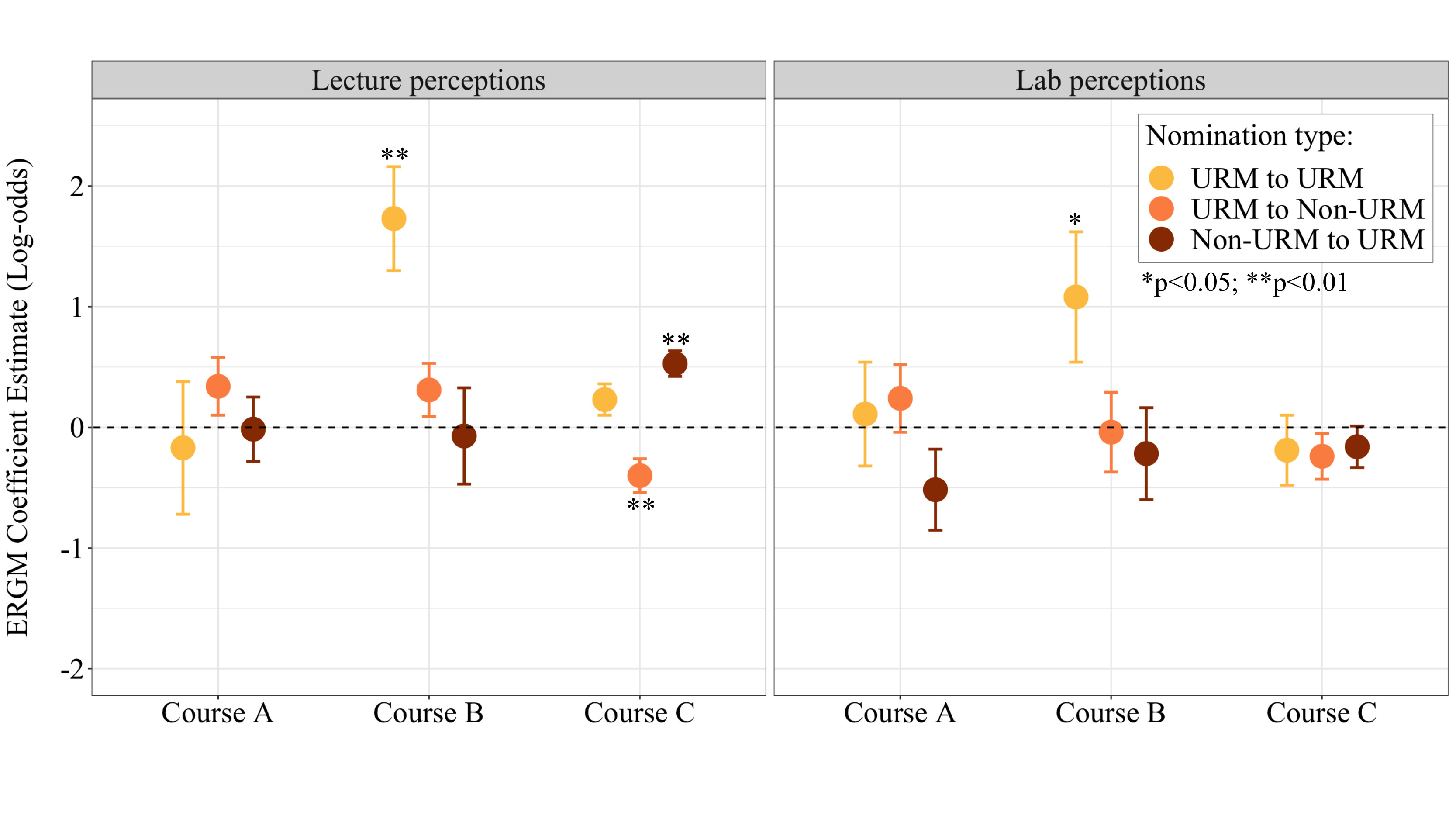}
\caption{Plot of ERGM coefficient estimates for the URM nomination variables. The base term (i.e., coefficient estimate of zero) is nominations from non-URM to non-URM. Error bars are the standard errors of the coefficient estimates (values shown in Table \ref{tab:newmodel}) and asterisks indicate statistical significance. 
}
\label{urmcoef}
\end{figure*}

The gender patterns suggested by our model fits are illustrated in Figs. \ref{nonlabgendernetworks} and \ref{labgendernetworks}. Despite there being no clear celebrities in the lecture perception network of Course A (the top-nominated man and woman received five and four nominations, respectively), our statistical model indicates a gender bias in this network. Therefore, men, on average, receive more nominations than women (the average size of the blue nodes is greater than the average size of the yellow nodes). For the lab perception network of Course A, men and women have an even distribution of nominations (the average size of the blue nodes is similar to the average size of the yellow nodes) as indicated by our statistical analysis. On the other hand, all three emergent celebrities in Course B's lecture perception network are men (the three largest  nodes, outlined in bold, are blue). In this network, the top-nominated man received ten times as many nominations (30) as the top-nominated woman (three). Course B's lab perception network is less centralized around a few celebrities, but we see that, on average, men received more nominations than women (the average size of the blue nodes is greater than the average size of the yellow nodes). For Course C, the lecture perception network has one male celebrity and a second celebrity with unknown gender (these two nodes are outlined in bold and overlap in the diagram). As mentioned above, the latter might explain why we did not resolve any gender bias in our model fit. Finally, Course C's lab perception network is similar to that of Course A in that there is a relatively even distribution of nominations across men and women, in line with our quantitative findings above.

We examine the coefficient estimates for the race/ethnicity variables in a similar manner. These results are summarized in Fig. \ref{urmcoef}. In both networks of Course A and the lab perception network of Course C, none of the coefficient estimates for the race/ethnicity variables are statistically significant after adjusting for the remaining variables in the model. This means that no particular nomination type (e.g., URM student nominating URM student) occurs more frequently than another -- both URM and non-URM students proportionately nominate their URM and non-URM peers (no racial/ethnic bias). In both networks of Course B, however, URM students disproportionately over-nominate URM peers, even after controlling for the other network configurations in the model (yellow dots in Fig. \ref{urmcoef}). Similarly, in Course C's lecture perception network, URM students significantly under-nominate non-URM peers (orange dots in Fig. \ref{urmcoef}) and non-URM students disproportionately over-nominate URM peers (brown dots in Fig. \ref{urmcoef}). Accordingly, one of the two prominent celebrities in this particular network (shown in Fig. \ref{nonlabgendernetworks}) is a URM student. 

Similar to the results related to gender, we note that the results vary for the lecture perception network of Course C if we impute the race/ethnicity of the second celebrity whose race/ethnicity is unknown. If we impute this student as non-URM, we find that URM students disproportionately over-nominate URM peers in this network, with no change to the other two race/ethnicity variables. If we impute this student as URM, we find that both URM and non-URM students disproportionately over-nominate URM peers. In both cases, therefore, we still find a tendency for URM students to receive more nominations than their non-URM peers.



\subsection{Roles of final course grade, outspokenness, and section enrollment in shaping peer recognition}

The remaining predictor variables in the model lend insight into the association between final course grade, outspokenness on the online discussion forum, and section enrollment and the structure of our observed perception networks. All coefficient estimates for the \textit{final course grade of nominee} variable, summarized in Table \ref{tab:newmodel}, are positive and statistically significant. That is, in all three courses and in both contexts, students with higher final course grades receive significantly more nominations than students with lower final course grades. The magnitudes of the coefficients also suggest that final course grade is a stronger predictor of receiving nominations in the lecture context than the lab context in every course, though again such comparisons should be considered tentative~\cite{duxbury2021problem}. We provide plots of the indegree distributions by final course grade in the Supplementary Material.

With regard to outspokenness (the \textit{discussion forum contributions of nominee} variable, which was only measurable for Courses B and C), we find that students who contribute more to the discussion forum are significantly more likely to receive nominations as strong in the lecture context. Contributions to the discussion forum, however, are not a significant predictor of receiving nominations as strong in the lab context. We provide plots of the indegree distributions by number of discussion forum contributions in the Supplementary Material.

We observe similar patterns across courses regarding the relationship between lab and discussion section enrollment and recognition among peers. Coefficient estimates for the \textit{homophily on lab section} variable are positive and statistically significant in every observed network, meaning that students are more likely to nominate peers in their lab section than peers outside of their lab section as strong in both lecture and lab content. Viewing the magnitude of the coefficients~\cite{duxbury2021problem}, this effect is, unsurprisingly, more pronounced in lab perception networks than lecture perception networks. On the other hand, coefficient estimates for the \textit{homophily on discussion section} variable are positive and statistically significant in all three lecture perception networks, but they are not statistically significant in any of the three lab perception networks. This suggests that students tend to nominate peers in their discussion section as strong in the lecture material, but they do not systematically nominate discussion peers as strong in the lab material.

\section{Discussion}

In this study, we collected students' nominations of strong peers in three different remote, introductory physics courses with varying student populations. We advance previous work by measuring both gender and racial/ethnic biases and differentiating perceptions related to lecture and lab contexts. The remainder of this section synthesizes our findings for each research question and concludes by noting recommendations for instruction and limitations to the study.

\subsection{Mixed evidence of gender and racial/ethnic biases in recognition}

Our analyses found mixed results regarding the presence or absence of gender bias in students' recognition of their peers. After adjustments for various measures reflecting structural tendencies of tie formation, women proportionately nominated their male and female peers in all courses and contexts (lecture and lab) except Course C's lecture perception network. In this network, women disproportionately nominated other women over men as strong in the lecture material. In contrast, men proportionately nominated their male and female peers in three out of six observed networks (lab perception network of Course A and both networks of Course C) after controlling for other network configurations. Men significantly under-nominated their female peers in Course A's lecture perception network and in both perception networks of Course B. Recall that if we impute the second celebrity in Course C's lecture perception network as a man, men also significantly under-nominated their female peers in this network.

The results related to gender bias in lecture peer perceptions add insight to those found in prior work~\cite{grunspan2016,salehi2019,bloodhart2020}. Across these studies, the courses vary by student population (majors, course level, and gender), instructional type (traditional and interactive lectures or non-traditional labs), and institution. This variability, understandably, leads to different conclusions in each study (including across the courses examined in our study). Contrary to expectations, a course's gender composition (whether gender-balanced, majority men, or majority women) and discipline (whether physical sciences and engineering or biology) do not seem to predict the presence or absence of a gender bias in students' recognition of their peers. Neither does the instructional style -- whether traditional lecture, interactive lecture, or lab instruction -- or class size (whether the course contains 90 or 400 enrolled students). 

One common factor that \textit{is} consistently associated with gender bias across different studies, however, is the course level. Across the four studies, courses at the first-year level (those in Ref.~\cite{grunspan2016}, Courses A and B in our study, and those in Ref.~\cite{bloodhart2020}, assuming the student populations in the lower level courses are primarily first-year based on typical course enrollments) all exhibit gender biases, whereas those at the beyond-first-year level (those in Ref.~\cite{salehi2019} and Course C in our study) do not. We posit that developing familiarity and friendship with peers in previous semesters allows for a more diverse set of students to gain recognition in subsequent courses. Students in beyond-first-year courses, for example, likely have had more opportunities to showcase their knowledge or skills in front of their peers during prior courses they take together. In students' first introductory courses, in contrast, a gender bias in peer recognition aligned with sociohistorical stereotypes~\cite{danielsson2012exploring,gonsalves2016masculinities,kessels2006goes,schmader2004,makarova2015trapped,makarova2019} endures before the students get to know each other. Alternatively, this pattern could be due to selection effects where only those women who received substantial recognition in their first course enrolled in subsequent courses. Thus, all of the gender bias may have occurred in the first year courses, creating unequal representation of students in the subsequent courses where we no longer find a bias. We note that this relationship between course level and gender bias in peer recognition is a tentative claim given the celebrity of unknown gender in Course C's lecture perception network. 

The modified analyses used in our study also add to the understanding of the nature of the gender bias in introductory STEM courses. As in the perceptions study in biology~\cite{grunspan2016}, we found that, when a gender bias in peer recognition existed, men under-nominated women, but women proportionately nominated both men and women. This result differs from that of Bloodhart and colleagues~\cite{bloodhart2020}, however, who found that \textit{both} men and women tend to under-nominate women. More details about their analyzed courses are necessary to determine which, if any, course features may have led to these different results.

Our study also uniquely evaluated whether a racial/ethnic bias exists in students' recognition of peers. Race/ethnicity was not a significant predictor of nominations in either network of Course A or the lab perception network of Course C. In both networks of Course B and the lecture perception network of Course C (even when imputing the second celebrity of the latter network as a URM or non-URM student), however, URM students were more likely to receive nominations than their non-URM peers. This suggests that, when the nomination probabilities are adjusted for other variables in the model, URM students received more recognition than their non-URM counterparts, despite the documented stereotypes against URM students in science~\cite{tate2005does,ceglie2011underrepresentation,grossman2014perceived,blaine2013understanding,eaton2020,carlone2007understanding} and indications that URM students report significantly lower senses of recognition than non-URM peers in their physics courses~\cite{lock2013physics,kalender2019gendered,hazari2013science}. 

We discuss several possible explanations that may have influenced these surprising findings. First, for the networks where we found no racial/ethnic bias (both networks of Course A and the lab perception network of Course C), one might expect low statistical power would explain the lack of a measurable effect: URM students made up less than 30\% of each analyzed course. This explanation is unlikely, however, given that we were able to statistically discern an effect in the other networks with comparable proportions of URM and non-URM students. Alternatively, we note that this study was fielded in the aftermath of the Black Lives Matter protests following the murder of George Floyd. Students (especially at Cornell University) were aware of the political climate~\cite{floyd1,floyd2}, which might have created more awareness of racial/ethnic bias (compared to gender bias) and thus social desirability biases in the responses. We note the plausibility of this explanation given that the effect on URM students' nominations is either unbiased or in the \emph{opposite} direction to what research would have predicted. Another explanation, particularly for the tendency for URM students to nominate URM peers in Course B, is friendship tie homophily. A host of research suggests that friendship serves as a mechanism for recognition~\cite{alaee2022,laninga2018moderating} and that students tend to form friendships with peers of their same race/ethnicity~\cite{mcpherson2001birds,currarini2010identifying}. URM students in Course B, therefore, might have formed friendships with one another and in turn recognized one another as strong in the course. Finally, in this study we measured actual recognition, whereas some prior work measures perceived recognition~\cite{lock2013physics,kalender2019gendered,hazari2013science}. Students' actual recognition may differ from their perceptions of recognition, resulting in these different outcomes. For example, students from underrepresented groups may perceive lower recognition than they are actually given based on their awareness of existing stereotypes. We recommend for future research to directly compare perceived and actual recognition across demographic groups to examine this viable phenomenon.

\subsection{Recognition differs between lecture and lab contexts}


Different from previous studies, we probed peer perceptions related to lecture and lab contexts separately. We observed very different network structures across these contexts, with celebrities emerging in two out of three lecture perception networks but in none of the lab perception networks. We suspect that the structure of coursework in each context impacted the distribution of nominations. In the courses we analyzed, lectures contained half or all of the enrolled students (depending on whether there were one or two lecture sections). The few highly motivated students (i.e., the celebrities) who frequently participated in lecture by answering or asking questions in front of the rest of the class likely gained considerable recognition from peers as strong in the lecture context. Because lectures were held on Zoom, students could also readily see the names of these celebrities and recall them on the survey. A similar phenomenon may have also occurred during online office hours, which were (anecdotally) very busy. By contrast, labs were held on Zoom and used breakout rooms, so students were restricted to interacting with and seeing the names of just a few peers. Lab groups were also held stable throughout the semester, allowing for meaningful yet limited recognition~\cite{doucette2020there,hazari2018towards}. We note that while discussion sections also used online breakout rooms, they did not necessitate interaction between students (students submitted individual work, if at all, and some groups would leave their cameras and microphones off and work independently). Labs, on the other hand, required students to set up their experiments, collect and analyze data, and coordinate lab notes for a weekly group submission, all of which were negotiated through conversations.



Our findings pertaining to gender, moreover, suggest that students perceive male and female peers' expertise with lecture and lab material differently. The presence or absence of a gender bias in peer recognition varied between contexts in Course A, with a gender bias in lecture but not in lab perceptions. In Course B, there was a stronger gender bias in lecture perceptions than lab perceptions. The lecture, but not lab, perception results mostly agree with prior work, which has found a gender bias in course-level perceptions~\cite{grunspan2016,bloodhart2020}. We speculate that, as was the case in the study of Grunspan and colleagues~\cite{grunspan2016}, men may have been more verbally outspoken than women in the lecture sections (though we did not measure this). Indeed, it has been shown in introductory science courses that women are less comfortable participating in whole-class discussions than men~\cite{eddy2015caution} and that women respond less frequently than men to instructor-posed questions to the class~\cite{eddy2014,aguillon2020gender}. In remote courses in particular, research has found that men participate more than women both verbally and in the chat window and that students acknowledge chat messages from male peers more than female peers~\cite{nichols2022participation}. 
Recognition in labs, on the other hand, occurs among students working together in their lab groups. In our observed courses, lab groups were created based on a group-forming survey where students could indicate their preferences related to group work, for instance if they wished to work with (or not work with) certain peers. Instructors intentionally created lab groups based on the survey and also avoided groups with isolated women. One possible explanation for observing less gender bias in labs, therefore, is that women had sufficient opportunities for recognition within their majority-women or all-women lab groups. Alternatively, students may hold different conceptions of what it means to be ``strong in the lab/discussion section material" and ``strong in the lab material." This seems plausible given that we found more gender bias in the lecture context than the lab context, yet stereotypes typically associate physics and masculinity with technical skills (lab) \textit{and} natural brilliance (lecture)~\cite{danielsson2012exploring,gonsalves2016masculinities,kessels2006goes,schmader2004,makarova2015trapped,makarova2019}. To explore this possibility, we have modified the perceptions survey to also ask students to briefly explain their nominations. We will use these data to unpack the traits or behaviors students attribute to being strong in each context and whether this explains the difference in gender bias.

We also found that higher final course grades predict more received nominations from peers across all courses and contexts, in agreement with prior work~\cite{grunspan2016,salehi2019}. Results related to discussion forum contributions and section enrollment, however, varied by context. We found that students' contributions to the course discussion forum was a significant predictor of receiving nominations in lecture, but not in lab, perception networks. We suspect that this difference occurred because most content posted on the discussion forum was related to lecture material rather than lab material. It appears that participation in the discussion forum served as a means of becoming more visible to and recognized by peers, but only in regard to the content being discussed. The lecture perception networks, moreover, contained emergent celebrities in the two courses using a discussion forum, Courses B and C, but not in Course A. Frequent visibility in the discussion forum, where students' names are explicitly tied to their contributions, might be a mechanism for students receiving many nominations as strong in their course. Similar to previous studies~\cite{grunspan2016,salehi2019}, we also observed that lab section enrollment is important for shaping peer perceptions in both contexts, while discussion section enrollment is only a strong predictor of lecture perceptions. In other words, students learn about one another's strengths and acquire recognition related to lecture material in both discussion and lab sections, while they learn about each other's strengths related to lab material only in lab section.

\subsection{Recommendations for instruction}

Together, our results related to gender and racial/ethnic bias in peer recognition point to courses (first-year level) and contexts (both lecture and lab) in which students from underrepresented backgrounds (mostly women) may receive less recognition than their peers and, therefore, may be at a disadvantage for developing their physics identity~\cite{hazari2018towards, kalender2019female}. Because recognition is one of the most important dimensions of physics identity~\cite{hyater2018critical,lock2013physics,carlone2007understanding,hazari2010connecting,hazari2018towards,hazari2017importance,kalender2019gendered,godwin2016identity,godwin2017pushing}, instructors may support all students' identity development by facilitating more equitable peer recognition. 

Our findings suggest that instructors, particularly of STEM courses at the first-year level, should aim to create opportunities for meaningful recognition in all aspects of a course. For example, research suggests that friendship and collaboration with peers is one mechanism for recognition: interacting with others allows for students to showcase their knowledge and skills and acquire validation from others~\cite{alaee2022,quan2022trajectory}. Opportunities for recognition, therefore, may be achieved through more student-centered instructional styles, where students work closely with one another in small groups~\cite{hazari2018towards}. Though small group work is already common in labs, lectures often place emphasis on individuals answering questions in front of the class. Poll questions and other active learning activities implemented in lectures may be turned into group discussions and group submissions rather than individual work. Further, if students are presenting group work in front of the whole class, instructors can create opportunities for positive recognition by allowing groups to discuss the ideas before asking them to share, increasing the likelihood of groups landing on the correct answer~\cite{smith2009peer}. 

In terms of forming groups, prior research suggests that historically underrepresented students (women and URM students) benefit from working in groups where they are not isolated~\cite{heller1992teaching,sullivan2018small}. One study, for example, found that gender homogeneous and majority-women groups performed better when solving physics problems than majority-men groups~\cite{heller1992teaching}. The researchers observed in these majority-men groups that the men dominated the conversation, ignoring suggestions from their female peer. Our results agree with this work from a different perspective, namely recognition. We found that the gender bias in peer recognition was weaker in the lab context (where groups intentionally avoided isolated women) than the lecture context (where any group work was completed in unintentionally formed groups). Our results, therefore, support the prior work recommending that underrepresented students (based on both gender and race/ethnicity) be placed in groups where they are not the minority. Previous research also suggests that students tend to become friends with, and therefore may be more prone to recognize, peers of similar academic achievement~\cite{laninga2018moderating}. Intentionally forming heterogeneous groups of students based on performance, therefore, might enable students of different levels of academic achievement to recognize each other.

As to whether groups should be held the same or changed throughout the semester, our study cannot make a strong recommendation one way or another. Close and consistent collaboration with group members seems to allow students to overcome any implicit biases with which they enter the course~\cite{doucette2020there}. Meanwhile, altering groups may be beneficial in allowing students to gain recognition from many of their peers.   

Outside of group work, instructor-posed questions to the whole class can still provide many students with opportunities to gain recognition. For a given question, an instructor may ask for multiple volunteers to share their thinking before hearing from any individual student (``many hands'') or to randomly select individuals to share (``warm'' or cold call)~\cite{tanner2013structure}. The instructor can also explicitly ask for different volunteers each time. Instructors should also be cautious when allotting praise to students' answers to the whole class. If a student's answer is followed by ``Perfect!" there is little room for other students to contribute or ask questions, limiting exposure to peers to just the one student who volunteered. 

\subsection{Limitations}

We end this section by acknowledging multiple limitations of our study. First and foremost, we collected our data during a global pandemic. Students' learning experiences were certainly impacted by this event~\cite{wilcox2020recommendations,dew2021student,klein2021studying,marzoli2021,doucette2021newtothis} and, as a result, our findings may not be generalizable to physics instruction during normal circumstances. In addition, the courses analyzed here were almost entirely held online. While our results align with some of the previous work from in-person courses, future work should perform a more systematic comparison of peer perceptions in face-to-face and remote courses. 

With regard to our methods, our perceptions survey may not have captured all nominations. We did not provide students with a list of names to look at when filling out the perceptions survey, so students may not have remembered the names of individuals they perceived as strong in the material. Additionally, we only collected survey responses at the midpoint of the semester. Other work administered surveys either both at the midpoint and endpoint of the course~\cite{grunspan2016} or just at the endpoint of the course~\cite{salehi2019}. Future work comparing physics students' perceptions at multiple points in the course or just at the end of the course may add nuance to our findings.

We also performed text processing on the surveys to match the reported names to the class roster. This process dropped fewer than 5\% of the reported nominations (for instance, due to students misspelling a peer's name). We note the possibility for bias in the text processing because certain kinds of names might be less likely to be matched and thus more likely to be dropped from analysis. On one hand, students with common first or last names in the class might be less likely to be matched, particularly if only their first or last name is listed. Complicated names might be more prone to misspelling and, therefore, may also have a low probability of being matched. Rare first or last names, on the other hand, are more likely to be matched because they are unique. Whether certain kinds of students, for instance non-URM and URM students, have common or rare names might influence the representation of demographic groups in the data. Because we were able to match a high percentage of the data (more than 95\%), however, we do not believe this potential bias impacted our study's results. 

In addition, we categorized race/ethnicity in terms of URM status because the number of students in each racial/ethnic group was too small for our quantitative analysis to produce useful and interpretable results. However, this inevitably masks differences in recognition between students of individual racial/ethnic groups. Future research should seek to study more diverse student populations with statistically sufficient representation from all racial/ethnic groups. We also treated gender and race/ethnicity separately. It seems valuable for future work to determine whether gender and race/ethnicity are separately important for peer perceptions or whether it is the \textit{intersection} between gender and race/ethnicity that significantly explains recognition patterns. 

Finally, peer recognition might depend on other variables that we did not measure or analyze in this study. For example, research suggests that students view their friends as strong in the course~\cite{alaee2022,laninga2018moderating}. To determine whether this is the case, students' friendship ties could be collected and added as a predictor variable in the statistical model. Students' majors might also be related to recognition: perhaps students view peers in particular majors (e.g., STEM majors or physics majors in particular) as stronger in physics than peers in other majors (e.g., life sciences or non-STEM majors). We did not examine the relationship between student majors and recognition in this study because the courses were quite homogeneous on major (either most students were studying engineering or most students were studying physics), many students did not report their sub-field within the engineering school, and some students did not yet declare a major. Future work should investigate whether and how friendship ties, students' majors, and other variables relate to students' perceptions of strong peers.

\section{Conclusion}

Examining students' nominations of strong peers, we found variation in gender and racial/ethnic biases across three different remote, introductory physics courses. We observed that courses primarily serving first-year students exhibited a gender bias in lecture perceptions, while those serving beyond-first-year students did not. Additionally, URM students were either more or equally likely to receive nominations than their non-URM peers, contrary to what prior research would predict. Recognition also varied between lecture and lab contexts. Lecture perception networks contained a few central students receiving many nominations, however no outstanding celebrities emerged in the lab perception networks. These results suggest that recognition varies within different student populations and instructional contexts. Findings also point to advantages of instruction that emphasizes small group work and allows for many different students to speak up in front of the class. These instructional efforts are hopefully the first steps toward creating more widespread, rather than skewed, distributions of peer recognition, such that all students can develop their physics identities – a critical predictor of participation, persistence, and career intentions in physics and other science disciplines.

\section*{ACKNOWLEDGEMENTS}

This material is based upon work supported by the National Science Foundation Graduate Research Fellowship Program Grant No. DGE-2139899 and National Science Foundation Grant No. DUE-1836617. The authors thank Cole Walsh, David Wu, Andy Schang, and David Esparza for providing meaningful feedback on this work.

\bibliography{perceptions.bib} 

\section{Appendix}

\subsection{Original model results and goodness-of-fit comparison}

\begin{figure*}[t]
\includegraphics[width=6.5in,trim={0 3cm 0 0}]{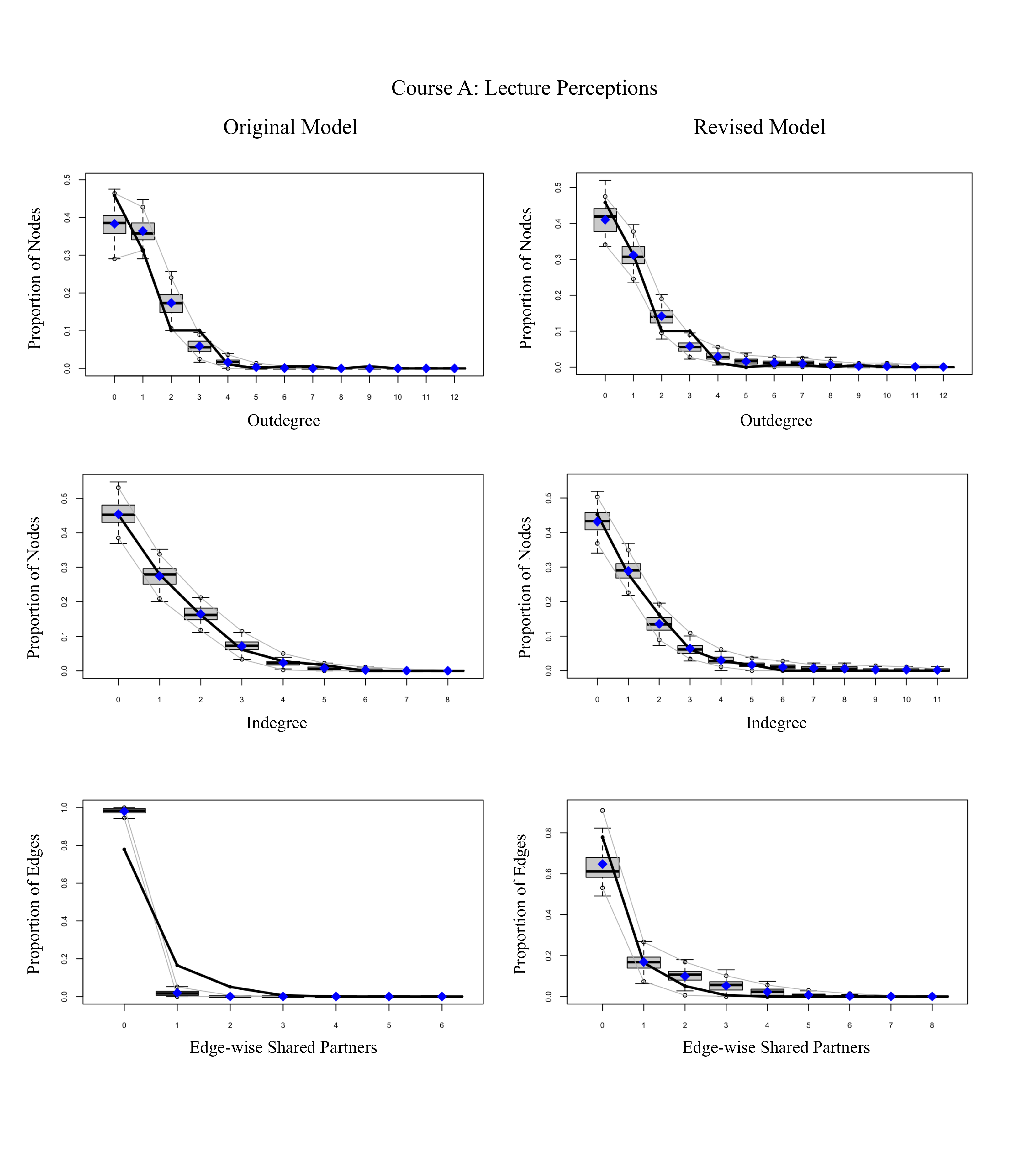}
\caption{Goodness-of-fit metrics for both the original ERGM model~\cite{grunspan2016} and the revised ERGM model for the lecture perception network of Course A.}
\label{gofnonlab}
\end{figure*}

Table \ref{tab:grunspanreplication} shows the coefficient estimates using the original model from Ref.~\cite{grunspan2016} with our data. We include four additional variables to measure discussion section homophily and racial/ethnic bias. Figure \ref{gofnonlab} compares the goodness-of-fit metrics of the original model and our revised model (presented in the main text). The horizontal axis represents a network measure -- outdegree (number of nominations made), indegree (number of nominations received), and edge-wise shared partners (measure of triadic closure) -- and the vertical axis represents the proportion of students who display that network measure. Plots show the distribution of the network measures in the observed data (thick black line) and the distribution for 10 network simulations from the model (boxplots). While the original model captures the outdegree and indegree distributions sufficiently, the revised model significantly improves how well the model captures the distribution of edge-wise shared partners.

\begin{table*}[!ht] 
\centering
  \caption{\label{tab:grunspanreplication} ERGM coefficient estimates using the original model of Grunspan and colleagues~\cite{grunspan2016}. Standard errors of the coefficient estimates are in parentheses. Asterisks indicate statistical significance ($^{*}$p$<$0.05; $^{**}$p$<$0.01). For every network, the goodness-of-fit metrics for the original model are better than those for the revised model (shown in Fig. \ref{gofnonlab}).} 
  \begin{ruledtabular}
\begin{tabular}{lcccccc} 
\\[-1.8ex] & \multicolumn{2}{c}{Course A} &  \multicolumn{2}{c}{Course B} & \multicolumn{2}{c}{Course C} \\ 
\cline{2-3}
\cline{4-5}
\cline{6-7}
\\[-1.8ex] & \multicolumn{1}{c}{Lecture} & \multicolumn{1}{c}{Lab} & \multicolumn{1}{c}{Lecture} & \multicolumn{1}{c}{Lab} & \multicolumn{1}{c}{Lecture} & \multicolumn{1}{c}{Lab} \\ 
\hline
 \textit{Edges} & -9.00$^{**}$ & -8.12$^{**}$ & -13.90$^{**}$ & -9.07$^{**}$ & -7.69$^{**}$ & -9.79$^{**}$ \\ 
  & (0.76) & (0.63) & (1.57) & (1.04) & (0.45) & (0.56) \\ 
  & & & & & & \\ 
 \textit{Mutuality} & 3.38$^{**}$ & 2.50$^{**}$ & 2.23$^{**}$ & 1.76$^{**}$ & 4.26$^{**}$ & 2.54$^{**}$ \\ 
  & (0.32) & (0.29) & (0.36) & (0.43) & (0.29) & (0.24) \\ 
  & & & & & & \\ 
   \textit{Isolates (0-indegree)} & 0.53$^{*}$ & 0.23 & 0.68 & 0.03 & 1.02$^{**}$ & 0.02 \\ 
  & (0.25) & (0.24) & (0.49) & (0.38) & (0.18) & (0.19) \\ 
  & & & & & & \\ 
  \textit{Female nominator} & 0.20 & 0.14 & 0.65 & 0.41 & 0.29 & 0.16 \\ 
  & (0.24) & (0.25) & (0.42) & (0.45) & (0.18) & (0.21) \\ 
  & & & & & & \\ 
   \textit{Woman-woman bias} & 0.23 & 0.30 & -0.56 & 0.67 & 0.09 & 0.10 \\ 
  & (0.21) & (0.21) & (0.71) & (0.44) & (0.14) & (0.18) \\ 
  & & & & & & \\ 
 \textit{Man-man bias} & 0.52$^{*}$ & 0.48$^{*}$ & 1.01$^{**}$ & 0.70$^{*}$ & 0.307 & 0.52$^{**}$ \\ 
  & (0.21) & (0.20) & (0.37) & (0.35) & (0.16) & (0.178) \\ 
  & & & & & & \\ 
   \textit{URM nominator} & 0.06 & 0.14 & -0.37 & -0.01 & -0.61$^{**}$ & 0.11 \\ 
  & (0.25) & (0.25) & (0.31) & (0.46) & (0.21) & (0.21) \\ 
  & & & & & & \\ 
\textit{URM-URM bias} & -0.26 & -0.07 & 1.52$^{**}$ & 1.19 & 0.82$^{**}$ & -0.13 \\ 
  & (0.55) & (0.47) & (0.56) & (0.64) & (0.23) & (0.26) \\ 
  & & & & & & \\ 
\textit{Non-URM - non-URM} bias & 0.04 & -0.03 & -0.73$^{**}$ & 0.08 & -0.26$^{*}$ & 0.35$^{*}$ \\ 
  & (0.15) & (0.15) & (0.23) & (0.31) & (0.12) & (0.15) \\ 
  & & & & & & \\ 
  \textit{Final course grade of nominee} & 0.79$^{**}$ & 0.29 & 2.41$^{**}$ & 0.58$^{*}$ & 0.41$^{**}$ & 0.43$^{**}$ \\ 
  & (0.18) & (0.15) & (0.37) & (0.26) & (0.12) & (0.14) \\ 
  & & & & & & \\ 
 \textit{Discussion forum contributions} & N/A  & N/A  & 0.03$^{**}$ & 0.01$^{*}$ & 0.01 & 0.007 \\ 
 \textit{of nominee} &  &  & (0.004) & (0.006) & (0.006) & (0.009) \\ 
  & & & & & & \\  
  \textit{Homophily on lab section} & 1.20$^{**}$ & 3.68$^{**}$  & 0.64$^{**}$ & 3.20$^{**}$ & 1.13$^{**}$ & 4.37$^{**}$ \\ 
  &  (0.15) & (0.20)  & (0.18) & (0.34) & (0.15) & (0.16) \\ 
  & & & & & & \\ 
 \textit{Homophily on discussion section} & 1.84$^{**}$ & 0.14  & 0.61$^{**}$ & 0.13 & 1.77$^{**}$ & 0.46$^{*}$ \\ 
  &  (0.15) & (0.21)  & (0.19) & (0.24) & (0.14) & (0.18) \\ 
  & & & & & & \\ 
  \hline
\end{tabular} 
\end{ruledtabular}
\end{table*} 

\end{document}